\documentclass[runningheads]{llncs}
\usepackage[T1]{fontenc}
\usepackage{graphicx}
\usepackage{hyperref}

\usepackage{diagbox}
\usepackage{listings}
\usepackage{subcaption}

\begin{document}
\title{Representing and querying data tensors in RDF and SPARQL}
%
%
\author{Piotr~Marciniak\inst{1}\orcidID{0009-0009-4583-896X} 
\and Piotr~Sowiński\inst{1,2}\orcidID{0000-0002-2543-9461} 
\and Maria~Ganzha\inst{1,3}\orcidID{0000-0001-7714-4844}
}
\authorrunning{P. Marciniak et al.}
%
\institute{Warsaw University of Technology, pl. Politechniki 1, 00-661 Warsaw, Poland
\and
NeverBlink, ul. Wspólna 56, 00-684 Warsaw, Poland
\and
Systems Research Institute, Polish Academy of Sciences, \\ ul. Newelska 6, 01-447 Warsaw, Poland
}
\maketitle              

\begin{abstract}
Embedding tensors in databases has recently gained in significance, due to the rapid proliferation of machine learning methods (including LLMs) which produce embeddings in the form of tensors. To support emerging use cases hybridizing machine learning with knowledge graphs, a robust and efficient tensor representation scheme is needed. We introduce a novel approach for representing data tensors as literals in RDF, along with an extension of SPARQL implementing specialized functionalities for handling such literals. The extension includes 36 SPARQL functions and four aggregates. To support this approach, we provide a thoroughly tested, open-source implementation based on Apache Jena, along with an exemplary knowledge graph and query set.


\textbf{Resources:} \url{https://w3id.org/rdf-tensor}
\end{abstract}

\section{Introduction}

The increasing use of machine learning models (especially Large Language Models -- LLMs) has created a need for efficient storage and manipulation of tensor data, such as text embeddings. Consequently, a growing demand exists for solutions integrating data storage and querying with data tensor processing. Recently, several database extensions have been developed to address this, such as PostgreSQL with \textit{pgvector}, MongoDB with \textit{Atlas Vector Search}, or Redis with \textit{Redis VSS}. These extensions enable existing architectures to facilitate tensor processing (e.g., finding similar vectors based on cosine similarity). AllegroGraph\footnote{\url{https://allegrograph.com/products/neuro-symbolic-ai/}} is a solution from the RDF domain that integrates knowledge graphs and vectors. Unlike the earlier solutions mentioned, it uses an additional vector database for this purpose, only exposing semantic-level operations to users. However, a more tightly-integrated approach, with tensor operations happening directly in the RDF database, would be more general (covering more use cases) and on par with modern SQL and NoSQL database architectures~\cite{pan2024survey}.

A data tensor is a multidimensional array used to store and process data. Tensors generalize scalars, vectors, and matrices to higher dimensions. However, currently, there is no direct way to represent such high-dimensional arrays in RDF. RDF 1.1~\cite{wood2014concepts} does support containers and collections, allowing users to model structures where elements are organized using a dedicated vocabulary. While it is possible to model tensors within these structures, it significantly increases the memory footprint. Each element in a tensor must be explicitly represented by a triple, which is excessively verbose (Fig.~\ref{fig:tensors}). Furthermore, querying such representations in SPARQL is highly inefficient because it involves filtering elements of the lists using property paths. SPARQL lacks easy-to-use, built-in functions for performing complex operations on collections. This means that even simple tasks such as adding two vectors require manual filtering and joining appropriate elements of the collections.

\begin{figure}[h]
    \centering
    
    \begin{minipage}{0.45\textwidth}
        \centering
        \begin{lstlisting}[frame=single, basicstyle=\tiny]
:word :hasWordEmbedding :embedding0 .
:embedding0 rdf:type rdf:List ;
    rdf:first 1;
    rdf:rest :embedding1 .
:embedding1 rdf:first 2 ;
    rdf:rest :embedding2 .
:embedding2 rdf:first 3 ;
    rdf:rest rdf:nil .
        \end{lstlisting}
        \subcaption{Representation as \texttt{rdf:List}}
        
    \end{minipage}
    \hfill
    \begin{minipage}{0.45\textwidth}
        \centering
        \begin{lstlisting}[frame=single, basicstyle=\tiny]
:word :hasWordEmbedding :embedding0 .
:embedding0 rdf:type rdf:Seq ;
    rdf:_1 1 ;
    rdf:_2 2 ;
    rdf:_3 3 ;
        \end{lstlisting}
        \subcaption{Representation as \texttt{rdf:Seq}}
    \end{minipage}    
    \caption{Data tensor representations currently possible with pure RDF 1.1.}
    \label{fig:tensors}
\end{figure}

The recent proposal for RDF datatypes for lists and maps~\cite{list_maps} could be used to represent multidimensional arrays, i.e., data tensors, as lists of lists. However, this proposition lacks specific SPARQL functions and operators, which would be helpful in processing these types of data. Additionally, multidimensional tensors would have to be represented as ``zig-zag'' nested lists. This approach is much more error-prone than dedicated tensor datatypes, as it does not ensure consistent dimensions of the tensors.

In this work, we introduce a framework for representing and processing multidimensional data tensors in RDF and SPARQL. The proposed approach defines new RDF datatypes for tensors and extends SPARQL capabilities with a comprehensive set of functions for tensor manipulation. This work is accompanied by a public specification of the extension, ontology files for datatypes and SPARQL functions, and a complete open-source implementation on top of Apache Jena: \textbf{\url{https://w3id.org/rdf-tensor}}

\section{Approach}

In the proposed approach, two new datatypes are introduced: \textbf{(i) numeric data tensor} for storing tensors containing numeric values (\texttt{dt:Numeric\-DataTensor}); \textbf{(ii) boolean data tensor}  (\texttt{dt:BooleanDataTensor}).

The lexical form for both datatypes follows the JSON specification~\cite{rfc8259}. In other words, a tensor is represented by a JSON string, which was chosen due to its readability and the wide availability of JSON libraries. Each representation consists of two mandatory keys: \texttt{data} and \texttt{shape}. The \texttt{data} key must contain an array of values, where all elements are either numbers or booleans, depending on the tensor type. The \texttt{shape} key must contain an array of integers, and the product of its elements must be equal to the length of the \texttt{data} array. The elements in the \texttt{data} array are assumed to be stored in row-major order (C-order), following the specified shape. An example of a two-dimensional boolean data tensor is presented in the following triple.

\begin{lstlisting}[frame=single, label={lst:boolean_data_tensors}, basicstyle=\tiny, language=SPARQL]
PREFIX dt: <https://w3id.org/rdf-tensor/datatypes#>

:s :p "{\"shape\":[1, 3],\"data\":[true, false, true]}"^^dt:BooleanDataTensor .
\end{lstlisting}

An additional key, \texttt{type}, is required for numeric data tensors. This key specifies the numeric type to ensure correct interpretation by an RDF processor. It can take one of six string values: three for real numbers (\texttt{float16}, \texttt{float32}, \texttt{float64}) and three for integers (\texttt{int16}, \texttt{int32}, \texttt{int64}). Example triples including a \texttt{dt:NumericDataTensor} are shown below.

\begin{lstlisting}[frame=single, label={lst:numeric_data_tensors}, basicstyle=\tiny,  breaklines=true, postbreak=\mbox{{$\hookrightarrow$}\space}, language=SPARQL]
PREFIX dt: <https://w3id.org/rdf-tensor/datatypes#>

:s :p1 
  "{\"type\":\"int32\",\"shape\":[2],\"data\":[1, 2]}"^^dt:NumericDataTensor .
:s :p2 
  "{\"type\":\"float32\",\"shape\":[2],\"data\":[1e0, 2.0]}"^^dt:NumericDataTensor .
\end{lstlisting}

The proposed solution also includes SPARQL functions and aggregates, which facilitate tensor manipulation. A total of 36 functions are defined that enable the transformation, comparison, and concatenation of tensors. These functions are categorized into six groups:

\begin{itemize}
    \item \textbf{Transformation functions}, which apply elementwise operations. For example, \texttt{dtf:cos} computes the cosine of each element in a numeric tensor.
    \item \textbf{Operators}, which perform arithmetic and logical computations with broadcasting. For instance, \texttt{dtf:add} adds two tensors element-wise, while \texttt{dtf:or} performs a logical OR operation on two Boolean tensors.
    \item \textbf{Indexing functions}, which enable tensor slicing. The function \texttt{dtf:getSubDT} extracts a sub-tensor based on a selection tensor.
    \item \textbf{Concatenation functions}, which merge tensors along a specified axis. For example, \texttt{dtf:concat} concatenates two tensors along a given dimension.
    \item \textbf{Reduction functions}, which aggregate tensor values along dimensions. The function \texttt{dtf:sum} computes the sum of tensor elements along a specified axis.
    \item \textbf{Similarity functions}, measuring distances between tensors. For instance, \texttt{dtf:euclideanDistance} calculates the Euclidean distance between tensors.
\end{itemize}

Consider the query below as an example of how the functions can be used. This query uses 3 functions in the \texttt{BIND} clause. \texttt{dtf:cosineSimilarity} calculates the cosine similarity of the two input tensors (literals with datatype \texttt{dt:NumericDataTensor}) and returns \texttt{xsd:double}. \texttt{dtf:norm1} calculates the first norm along the 0th dimension (first argument; dimensions are indexed from 0) of the difference tensor \texttt{dtf:minus} between two variables \texttt{?dt1} and \texttt{?dt2}. When run on the earlier presented triples, the query would assign to variable \texttt{?cos} the value $1$ and \texttt{?norm\_dt1} would have a value of 0 (both vectors are 1D so the result of a norm is a scalar), because both tensors have the same values.

\begin{lstlisting}[frame=single, label={lst:query}, basicstyle=\tiny,  breaklines=true, postbreak=\mbox{{$\hookrightarrow$}\space}, language=SPARQL]
PREFIX dtf: <https://w3id.org/rdf-tensor/functions#>

SELECT * WHERE {
    :s :p1 ?dt1
    :s :p2 ?dt2
    BIND(dtf:cosineSimilarity(?dt1, ?dt2) AS ?cos)
    BIND(dtf:norm1(0, dtf:minus(?dt1, ?dt2)) AS ?norm_dt1)
}
\end{lstlisting}

The proposed SPARQL extension also includes four new aggregates (\texttt{dta:sum}, \texttt{dta:avg},
\texttt{dta:var}, \texttt{dta:std}). The following query illustrates the usage of these aggregation functions. The query groups subjects and applies the \texttt{dta:sum} and \texttt{dta:avg} aggregates to compute the sum and average of tensors within each group. Since the aggregates work on numeric tensor literals, the resulting tensors maintain the highest numerical precision among the input numeric tensors (for numeric tensors presented earlier, the type would be \texttt{float32}).

\begin{lstlisting}[frame=single, label={lst:aggregate_query}, basicstyle=\tiny, breaklines=true, postbreak=\mbox{{$\hookrightarrow$}\space}, language=SPARQL]
PREFIX dta: <https://w3id.org/rdf-tensor/aggregates#>

SELECT ?s (dta:sum(?dt) AS ?sum_tensor) (dta:avg(?dt) AS ?avg_tensor) WHERE {
    ?s :p1|:p2 ?dt .
} GROUP BY ?s
\end{lstlisting}

\section{Implementation and results}

The presented approach was fully implemented for Apache Jena and released under the Apache 2.0 license\footnote{\url{https://github.com/RDF-tensor/jena-datatensor}}. The integration uses the Jena subsystem extension mechanism, so that it can be easily used with existing applications. This extension leverages the capabilities of the ND4J library which is a part of DeepLearning4J\footnote{\url{https://deeplearning4j.konduit.ai/}} to perform data tensor processing in a highly optimized manner, using vector capabilities of modern CPUs, or CUDA acceleration on Nvidia GPUs. The correctness of the extension was tested using unit and integration tests, configured in a similar manner to Apache Jena's tests for other components.

The provided repository also contains a directory with a converted subset of a scientific knowledge graph, including text embeddings~\cite{csgraph} and a set of exemplary queries showcasing the available functionalities.

\section{Conclusions}

By using the proposed data tensor datatypes, it is possible to process tensors in RDF much more efficiently than with pure RDF 1.1 constructs. Furthermore, the proposed SPARQL extension greatly simplifies processing such tensors with dedicated functions and aggregates. In the near future, we are hoping to gather feedback from interested parties on the design of the approach. The extension's specification should be then developed further.

In the current version, data tensors can only be represented using text. To accelerate the mapping from the lexical form to the value form, an efficient binary representation could be employed as the lexical form. This could be achieved by replacing the JSON string with, for example, a serialization of the same data in Protobuf, similar to the binary tensor formats used in Tensorflow and other machine learning frameworks~\cite{bogacka2024flexible,olston2017tensorflow}. Representing a binary-valued literal in text-based RDF formats like Turtle would however require using base64 encoding, which is an additional step slowing down serialization/deserialization and increasing file size. Therefore, we are currently exploring integrating raw binary tensor representations with the Jelly binary RDF format\footnote{\url{https://w3id.org/jelly}}~\cite{jelly}, through the use of embedded Protobuf messages, minimizing the serialization overhead.

\bibliographystyle{splncs04}
\bibliography{refs.bib}

\end{document}